\documentclass[aps,nofootinbib]{revtex4}
\usepackage{epsfig,graphics,amssymb,amsmath,subeqnarray}

\begin{document}

\title{Hydrodynamic friction of fakir-like  super-hydrophobic surfaces}
\author{Anthony M. J. Davis}
\author{Eric Lauga}
\email{Email: elauga@ucsd.edu}
\affiliation{Department of Mechanical and Aerospace Engineering, University of California San Diego, 9500 Gilman Drive, La Jolla CA 92093-0411, USA.}

\date{\today}
\begin{abstract}

A fluid droplet located on a super-hydrophobic  surface makes contact with the surface only at small isolated regions, and is mostly in contact with the surrounding air. As a result, a fluid in motion near such a  surface  experiences very low  friction, and super-hydrophobic surfaces display strong drag-reduction in the laminar regime.  Here we consider theoretically a super-hydrophobic surface composed of  circular posts (so called fakir geometry) located on a planar rectangular lattice.  Using a superposition of point forces with suitably spatially-dependent strength, we derive the  effective surface slip length for a planar shear flow on such a fakir surface as the solution to an infinite series of linear equations. In the asymptotic limit of small surface coverage by the posts, $\phi_s$, the series can be interpreted as Riemann sums,  and the slip length can be obtained analytically. 
For posts on a  square lattice of periodicity $L$, our analytical results predict that in the low $\phi_s$ limit, the surface slip length, $\lambda$, scales as 
$$\frac{\lambda}{L}\sim\frac{3}{16}\sqrt{\frac{\pi}{\phi_s}}
-\frac{3}{2\pi}\ln(1+\sqrt{2}),
$$
which is  in excellent quantitative agreement with previous numerical computations.
\end{abstract}
\maketitle

\section{Introduction}
 
\begin{figure}
\centering
\includegraphics[width=0.6\textwidth]{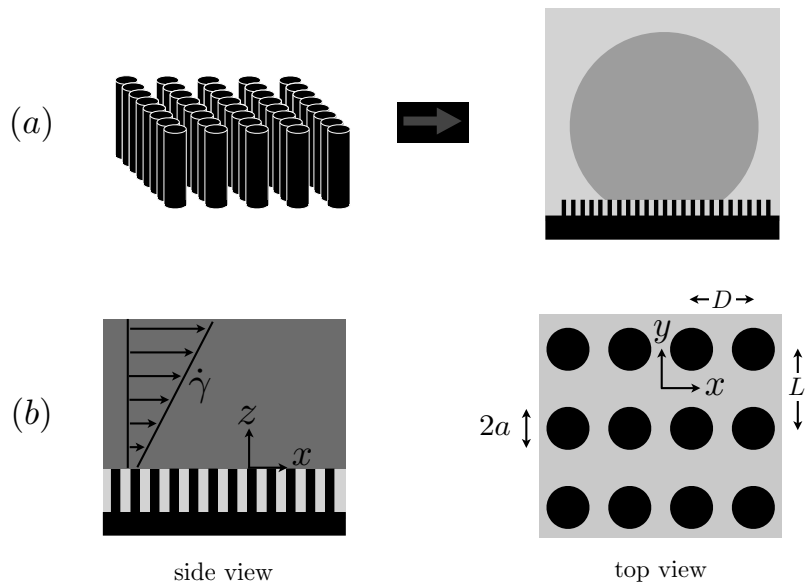}
\caption{Fakir-like super-hydrophobic surfaces  and setup for our calculation. 
(a): A droplet deposited on surface composed of tall hydrophobic posts can remain in a fakir super-hydrophobic state where it partially sits on the posts and partially on the air;
(b): Definition sketches. A viscous fluid of shear viscosity $\mu$ sits in a fakir state on a two-dimensional array of circular posts. The posts have radius $a$, and are arranged on a rectangular lattice of periodicity $D$ and $L$ along the $x$ and $y$ directions respectively. A shear flow with shear rate $\dot\gamma$ is imposed in the fluid along the $x$ direction, and perpendicularly to the surface. In-between the posts, the fluid-air interface is assumed to remain flat and parallel to the super-hydrophobic surface.}
\label{setup}
\end{figure}

When a small droplet of liquid is deposited on a hydrophobic surface whose geometry is sufficiently rough, in certain situations the liquid will not fill the roughness grooves but will instead adopt a lower energy configuration, sometimes called a fakir state, where it remains on top of the surface topography  (figure~\ref{setup}a). Exemplified by the Lotus leaf, surfaces for which this situation occurs are referred to as super-hydrophobic (or more generally, super-repellent), and their geometrical and physical description has been the subject of much recent work 
\citep{onda96,bico99, feng02,degennes_book,quere08,roach08}.

One of the remarkable properties of super-hydrophobic surfaces is their low friction opposing the flow \citep{rothstein_review}. Since the fluid next to such a surface makes contact with the solid at only a few isolated points and is mostly in contact with air, the shear stresses opposing fluid motion are small, and fluids in  fakir states can move very easily. In these situations, the surface friction is usually quantified by a slip length, $\lambda$, which is the  fictitious distance below the surface where the no-slip condition would be valid  on average \citep{neto05,bocquet07,laugareview}; no-slip corresponds to $\lambda=0$, while super-hydrophobic surfaces with $\lambda >0 $ have a lower effective friction than no-slip.

From a practical standpoint, many different methods exist to design and produce super-hydrophobic surfaces  \citep{feng02,roach08}, and they lead to surfaces with either random or controlled  geometrical features. Surfaces with random topography are known to show slip \citep{gogte05,choi06_PRL,joseph06}.  Planar surfaces with controlled geometry are of two kinds. The first kind are surfaces with  one-dimensional features such as long grooves, usually aligned  parallel or perpendicular to the flow direction, and for which a lot of experimental  \citep{ou04,ou05,truesdell06,choi06,maynes07,tsai09} and modeling  \citep{lauga03,cottin-bizonne04,davies06,maynes07,ng09,teo09} work has characterized their frictional properties.

The second kind are surfaces with two-dimensional patterning, usually a series of vertical posts distributed on a regular lattice, as illustrated schematically in figure~\ref{setup}a.  Only a handful of studies has focused on the friction of these types of surfaces, either experimentally  \citep{lee08} or theoretically  \citep{ybert07,ng09_2}. In the limit of low solid fraction, these two-dimensional surfaces are  expected however to display  much lower friction (algebraic divergence of the slip length) than one-dimensional surfaces (logarithmic divergence) \citep{ybert07}, and it is thus important  to be able to accurately predict their friction properties. This is the goal of the present paper.

Previous theoretical work focusing on surfaces composed of posts on a lattice found that if ${\phi_s} < 1$ denotes the areal density of the horizontal cross-section of the posts,  the effective slip length of the superhydrophobic surface, non-dimensionalized by the  typical distance, $L$, between the posts, behaves in the limit of low $\phi_s$ as
\begin{equation}\label{scaling}
\frac{\lambda}{L}\sim \frac{A}{\sqrt{\phi_s}} - B,
\end{equation}
\citep{ybert07,ng09_2}. In \eqref{scaling}, the positive coefficients, $A$ and $B$, depend on the lattice and posts geometry but not $\phi_s$, and can be fitted to numerical computations, while the square root scaling can be physically rationalized as follows. The shear stress acting on the posts scales as 
$\tau \sim \phi_s \mu \dot \gamma_s$ where $\mu$ is the fluid viscosity and $\gamma_s$ the typical shear rate around a post. If $U$ is the typical fluid velocity between the posts, and $a$ the typical post radius, we have $\dot \gamma_s \sim U/a$, and since $\lambda \sim \mu U/\tau$ we obtain 
$\lambda \sim a / \phi_s$. As the solid fraction scales as  
$\phi_s \sim (a/L)^2$, we have $a \sim L \phi_s^{1/2}$ and hence we obtain the square-root scaling of \eqref{scaling}, namely $\lambda / L \sim 1/\phi_s^{1/2}$ \citep{ybert07}.

In this work, we consider the simplest possible geometry for flow over a fakir-like super-hydrophobic surface, namely  vertical  posts with circular cross section located  on a regular rectangular lattice, and we provide two results. Firstly, we present an analytical method based on a linear superposition of flow singularities  to accurately determine the friction  (i.e. the effective surface slip length) for all surface coverage.  Secondly, by asymptotically considering the case of low solid fraction $\phi_s$, we mathematically derive the scaling coefficients $A$ and $B$ governing \eqref{scaling}, thereby predicting analytically the effective surface slip length. Our results are compared to previous computational work, and show  excellent agreement.

\section{Calculation of the effective surface  slip length}

\subsection{Problem outline}

Our geometry is illustrated in figure~\ref{setup}b. 
A viscous fluid of  viscosity $\mu$ is located  in a fakir state on a two-dimensional array of circular posts. The posts have radius $a$, and are arranged on a rectangular lattice of periodicity $D$ and $L$ along the $x$ and $y$ directions respectively, with $2a<\min(D,L)$.
A shear flow with shear rate $\dot\gamma$ is imposed in the far field by the prescribed velocity along the $x$ direction, with shear perpendicularly to the surface along the $z$ direction. The prescribed velocity is thus given by 
\begin{equation}\label{velinf}
{\bf v}\sim(\dot{\gamma}z+U){\bf e}_x \hbox{ as }z\to\infty,
\end{equation}
where ${\bf e}_x$, ${\bf e}_y$, ${\bf e}_z$ denote unit vectors 
parallel to the axes, 
and thus   $U$ denotes the effective slip velocity at the fakir contact plane, $z=0$.  In-between the posts, the fluid-air interface is assumed to remain flat and parallel to the underlying super-hydrophobic surface with no shear acting on it. The flow satisfies the no-slip boundary condition on the top of the circular posts. 

Our aim is to accurately estimate the effective slip length, that is, the mean speed/mean shear ratio at the super-hydrophobic plane.  As in our recent calculation for a moving grid \citep{davis09_2}, the basic tool 
here is the Stokes flow generated by a similarly periodic array of 
aligned, parallel point forces directed in their plane. In addition,  the force density function is assumed, for the leading  approximation, to have an inverse square root rim singularity, as in the creeping flow past a disk. The resulting linear system has matrix  elements that are interpreted as Riemann sums and the limit double  integrals are easily evaluated. The dominant error arises  from the missing term in the double sum and thus precise estimates can be 
obtained for the two coefficients in \eqref{scaling}.

\subsection{Calculation using a superposition of singularities}

The Reynolds number of the viscous incompressible flow is assumed to be
sufficiently small for the velocity field ${\bf v}$ to satisfy the
creeping flow (Stokes) equations \citep{happel}
\begin{equation}\label{stokes}
\mu\nabla^2{\bf v}=\nabla p, \qquad \nabla\cdot{\bf v}=0,  
\end{equation}
where $p$ is the dynamic 
pressure. 
The fluid motion can be represented as due to a distribution of
tangentially directed Stokeslets over the circular regions, augmented
by the uniform flow $U{\bf e}_x$. The density functions must be both 
periodic in two dimensions and symmetric with respect to the sides of 
each rectangle. The field due to a two-dimensional rectangular array, periods $D$
and $L$, of point forces of strength $4\pi\mu U\sqrt{DL}$ directed
parallel to the flow at infinity, is governed by
\begin{equation}\label{conty}
\nabla\cdot{\bf v}_A=0,
\end{equation} 
\begin{eqnarray}
\mu\nabla^2{\bf v}_A-\nabla p_A & =  & 4\pi\mu U\sqrt{DL}{\bf e}_x\delta(z)
\sum_{n_1=-\infty}^\infty\sum_{n_2=-\infty}^\infty\delta(x-n_1D)
\delta(y-n_2L) \\
& = & \frac{4\pi\mu U}{\sqrt{DL}}{\bf e}_x\delta(z)\sum_{m_1=-\infty}^\infty
\sum_{m_2=-\infty}^\infty\exp\left[2\pi i\left(\frac{m_1x}{D}
+\frac{m_2y}{L}\right)\right]\cdot     
\label{forcearray}\end{eqnarray}
The $m_1=0=m_2$ term in (\ref{forcearray}) yields the anticipated shear 
at infinity and, as in \citep{davis09_2}, the flow generated by the
oscillatory forcing is readily found by Fourier transform techniques \citep{hasimoto59}.
Thus the flow governed by (\ref{forcearray}) is compactly expressed as
\begin{equation}\label{arrayflow}
{\bf v_A}=U\left[\left(\frac{2\pi z}{\sqrt{DL}}-S_1\right){\bf e}_x+
\nabla\frac{\partial S_2}{\partial x}\right], \qquad 
p_A=\mu U\frac{\partial S_1}{\partial x},
\end{equation}
where, with a prime denoting that the $m_1=0=m_2$ term is missing,
\begin{eqnarray}
S_1&=&\frac{1}{\sqrt{DL}}\sum_{m_1}\sum_{m_2}{}' \frac{1}{\sqrt{(\frac{m_1}{D})^2
+(\frac{m_2}{L})^2}}  \nonumber
\\
&& \times \exp\left[2\pi\left(\frac{im_1x}{D}+\frac{im_2y}{L}
-z\sqrt{\left(\frac{m_1}{D}\right)^2+\left(\frac{m_2}{L}\right)^2}\right)\right],
\label{ess1}
\end{eqnarray}
and
\begin{eqnarray}
S_2&=&-\frac{1}{8\pi^2\sqrt{DL}}\sum_{m_1}\sum_{m_2}{}'\frac{1}
{(\frac{m_1}{D})^2+(\frac{m_2}{L})^2}\left(
\frac{1}{\sqrt{(\frac{m_1}{D})^2+(\frac{m_2}{L})^2}}+2\pi z\right) 
\nonumber
\\
&& \times\exp\left[2\pi\left(\frac{im_1x}{D}+\frac{im_2y}{L}
-z\sqrt{\left(\frac{m_1}{D}\right)^2+\left(\frac{m_2}{L}\right)^2}\right)\right],
\label{ess2}
\end{eqnarray}
with $S_1=\nabla^2 S_2$. 
Only the velocities at the screen and at infinity are needed for the
subsequent analysis.  The solution (\ref{arrayflow}) shows that
\begin{equation}\label{veeAinf}
{\bf v}_A\sim\frac{2\pi Uz}{\sqrt{DL}}{\bf e}_x \hbox{ as }z\to\infty,
\end{equation}
as well as 
\begin{eqnarray}
[{\bf v}_A]_{z=0} & = & U{\bf e}_x\left(-S_1+\frac{\partial^2S_2}
{\partial x^2}\right)_{z=0} \nonumber \\
&=&-\frac{U}{\sqrt{DL}}{\bf e}_x\sum_{m_1}\sum_{m_2}{}'
\frac{C(m_1,m_2)}{\sqrt{(\frac{m_1}{D})^2+(\frac{m_2}{L})^2}}\exp\left[2\pi i
\left(\frac{m_1x}{D}+\frac{m_2y}{L}\right)\right],
\label{veeAarray}
\end{eqnarray}
where, after substitution of (\ref{ess1}) and (\ref{ess2}), 
\begin{equation}\label{ceem}
C(m_1,m_2)=1-\frac{1}{2}\frac{(\frac{m_1}{D})^2}
{(\frac{m_1}{D})^2+(\frac{m_2}{L})^2}\cdot
\end{equation}

With suitably scaled force density functions in the circular regions, 
the total flow field is given by
\begin{equation}\label{vee}
{\bf v}=U{\bf e}_x+\int_{-\pi}^{\pi}\int_0^a
\left[\sum_{n=0}^\infty \frac{f_n(\alpha)}{\pi a^2}\cos 2n\beta\right]{\bf v}_A\left[r\cos\theta-\alpha\cos\beta,
r\sin\theta-\alpha\sin\beta,z\right]\alpha d\alpha d\beta,
\end{equation}
in which the symmetry conditions restrict the density function to even
cosines. The flow at infinity is determined by substitution of
(\ref{veeAinf}) into (\ref{vee}), which gives
\begin{equation}\label{flowinf}
{\bf v}\sim U{\bf e}_x+\frac{2Uz}{a^2\sqrt{DL}}{\bf e}_x
\int_{-\pi}^{\pi}\int_0^a\left[\sum_{n=0}^\infty f_n(\alpha)
\cos 2n\beta\right]\alpha d\alpha d\beta,\hbox{ as }z\to\infty.
\end{equation}
Comparison with (\ref{velinf}) then shows that the shear rate is given by
\begin{equation}\label{shear}
\dot{\gamma}=\frac{4\pi U}{a^2\sqrt{DL}}\int_0^a f_0(\alpha)
\alpha d\alpha .
\end{equation}
Thus only the mean force density contributes to the required slip
length, $\lambda$, given by
\begin{equation}\label{slip}
\lambda=\frac{\hbox{slip speed}}{\hbox{shear rate}}=\frac{U}
{\dot{\gamma}}=\frac{\sqrt{DL}}{4\pi}\frac{a^2}{\int_0^a f_0(\alpha)
\alpha d\alpha}.
\end{equation}

No-slip on the surface of the posts  is achieved by enforcing ${\bf v}={\bf 0}$ at $z=0,\,r<a$. The
substitution of (\ref{veeAarray}) into (\ref{vee}) gives the condition
\begin{eqnarray} 
\nonumber
1 &=& \frac{1}{\pi a^2\sqrt{DL}}\int_{-\pi}^{\pi}\int_0^a
\left[\sum_{n=0}^\infty f_n(\alpha)\cos 2n\beta\right]
\sum_{m_1}\sum_{m_2}{}'\frac{C(m_1,m_2)}{M}\\
&& \times\exp[2\pi iM\{r\cos(\theta-\psi)-\alpha\sin(\theta-\psi)\}]
\alpha d\alpha d\beta 
 \,  (0\leq r<a,-\pi<\theta\leq \pi).
\label{veezero1}
\end{eqnarray}
where we have defined
\begin{equation}\label{ellipt}
M(\cos\psi,\sin\psi) = \left(\frac{m_1}{D},\frac{m_2}{L}\right).  
\end{equation}
We next make two uses of the integral 
\begin{equation}\label{int1}
\frac{1}{\pi}\int_{-\pi}^{\pi}\cos 2n\beta \exp[\pm 2\pi iM\alpha
\cos(\beta-\psi)]d\beta=2(-1)^nJ_{2n}(2\pi M\alpha)\cos 2n\psi, 
\end{equation} 
where $J_p$ refers to the Bessel function of the first kind of order $p$,
first for a direct evaluation in (\ref{veezero1}) which gives,
\begin{eqnarray} 
\nonumber
1 & = & \frac{2}{a^2\sqrt{DL}}\int_0^a\sum_{n=0}^\infty f_n(\alpha)
(-1)^n
\sum_{m_1}\sum_{m_2}{}'
\frac{C(m_1,m_2)}{M}J_{2n}(2\pi M\alpha) \\
&& \times\exp[2\pi iMr\cos(\theta-\psi)]\cos 2n\psi
\alpha d\alpha  \qquad  (0\leq r<a,-\pi<\theta\leq \pi).
\label{veezero2}
\end{eqnarray}
and then to write down the Fourier coefficients in (\ref{veezero2}).
Thus we obtain 
\begin{eqnarray}
\delta_{k0} &= & \frac{2}{a^2\sqrt{DL}}\int_0^a\sum_{n=0}^\infty 
f_n(\alpha)
(-1)^{n-k}\sum_{m_1}\sum_{m_2}{}'\frac{C(m_1,m_2)}{M}J_{2n}(2\pi M\alpha)
\nonumber
\\
\label{veezero3}
&& \times J_{2k}(2\pi Mr)\cos 2k\psi\cos 2n\psi\,\alpha d\alpha  
\qquad (0\leq r<a,k\geq 0),
\end{eqnarray}
where $\delta_{kn}$ denotes the Kronecker delta.

We next make two uses of the integral 
\begin{equation}\label{int2}
\int_0^1\frac{x^{2n+1}J_{2n}(bx)}{\sqrt{1-x^2}}dx=\sqrt{\frac{\pi}{2b}} 
J_{2n+1/2}(b)  \qquad  (n\geq 0,b>0), 
\end{equation}
first by introducing the conveniently scaled, inverse square root 
approximations for the force density functions, 
\begin{equation}\label{densityfns}
f_n(\alpha)\sim\frac{\alpha^{2n}c_n}{a^{2n-1}\sqrt{a^2-\alpha^2}}
  \qquad  (n\geq 0), 
\end{equation}
whose substitution in (\ref{veezero3})  yields
\begin{eqnarray}
\delta_{k0} & = & \frac{1}{\sqrt{DL}}\sum_{n=0}^\infty c_n(-1)^{n-k}
\sum_{m_1}\sum_{m_2}{}'\frac{C(m_1,m_2)}{M\sqrt{Ma}}J_{2n+1/2}(2\pi Ma)
\nonumber
\\
&& \times 
J_{2k}(2\pi Mr)\cos 2k\psi\cos 2n\psi \qquad (0\leq r<a,k\geq 0).
\label{veezero4}
\end{eqnarray}
Then the immaterial $r$-dependence can be jettisoned by applying the 
operator,
$$\int_0^a \left(\frac{r}{a}\right)^{2k+1}\frac{dr}{\sqrt{a^2-r^2}}, $$
to (\ref{veezero4}), in order to obtain 
\begin{eqnarray}
\nonumber
\delta_{k0}  & = & \frac{1}{2a\sqrt{DL}}\sum_{n=0}^\infty c_n(-1)^{n-k}
\sum_{m_1}\sum_{m_2}{}'\frac{C(m_1,m_2)}{M^2}J_{2n+1/2}(2\pi Ma)\\
&& \times J_{2k+1/2}(2\pi Ma)\cos 2k\psi\cos 2n\psi  
\qquad (k\geq 0).
\label{linsys}
\end{eqnarray}
Equation~\eqref{linsys} is a symmetric infinite system of linear equations for the coefficients  $\{c_n;n\geq 0\}$. The substitution of (\ref{densityfns}) into (\ref{slip}) then leads to the slip length as
\begin{equation}\label{slip0}
\lambda=\frac{\sqrt{DL}}{4\pi c_0}\cdot
\end{equation}

\subsection{Asymptotic estimate of $\lambda$ in the low-$\phi_s$ limit}

An asymptotic estimate of the coefficients $\{c_n;n\geq 0\}$ is found
by noting that
\begin{equation}
\frac{a^2}{DL}\sum_{m_1}\sum_{m_2}{}'\frac{C(m_1,m_2)}{(Ma)^2}
J_{2n+1/2}(2\pi Ma)J_{2k+1/2}(2\pi Ma)\cos 2k\psi\cos 2n\psi,
\end{equation}
with
\begin{equation}
C(m_1,m_2)=1-\frac{1}{2}\cos^2\psi, \quad 
Ma=\sqrt{\left(\frac{m_1a}{D}\right)^2+\left(\frac{m_2a}{L}\right)^2},
\end{equation}
is a Riemann sum, with increments $(a/D,a/L)$ in $(x,y)$. In 
terms of polar coordinates $(r,\psi)$, the integral has the separable form
\begin{equation}
\int_{-\pi}^{\pi}\int_0^\infty\left(1-\frac{1}{2}\cos^2\psi\right)J_{2n+1/2}
(2\pi r)J_{2k+1/2}(2\pi r)\cos 2k\psi\cos 2n\psi\frac{dr}{r}d\psi =\frac{3\pi\delta_{nk}}{2\epsilon_n(4n+1)},
\label{RSint}
\end{equation}
where Neumann's symbol $\epsilon_0=1,\epsilon_n=2\,(n>0)$. The
corresponding solution of (\ref{linsys}) is
\begin{equation}\label{soln0}
c_0=\frac{4a}{3\pi\sqrt{DL}},\qquad c_n=0 \,(n>0),
\end{equation}
whose substitution into (\ref{slip}) yields the dimensionless slip
coefficient,
\begin{equation}\label{slipcoeff0}
\frac{\lambda}{\sqrt{DL}}\sim\frac{3\sqrt{DL}}{16a}=
\frac{3}{16}\sqrt{\frac{\pi}{\phi_s}},
\end{equation}
where $\phi_s=\pi a^2/DL$ is the fractional surface area covered by
the circular regions.  We note that this leading-order term is symmetric in $(D,L)$,  the respective periods
along and across the imposed shear flow, corresponding therefore to an isotropic surface friction at this order.

The leading error in the estimate (\ref{slipcoeff0}) is due to the 
absence of a term associated with $m_1=0=m_2$ in the Riemann sum. If 
the latter is regarded as having central function values, then the 
integrable singularity at the origin can be handled by exact 
integration over the rectangle centred at $(0,0)$, assisted by small 
argument estimates of the Bessel functions. Then (\ref{RSint}) has the 
additional term
$$-\frac{4\pi^{2n+2k+1}}{\Gamma(2n+3/2)\Gamma(2k+3/2)}\int_0^{a/2L}
\int_0^{a/2D}\left[1-\frac{x^2}{2(x^2+y^2)}\right](x^2+y^2)^{n+k-1/2}
dxdy $$
\begin{equation}\label{RSinterr}
=O\left[\left(\frac{a^2}{DL}\right)^{n+k+1/2}\right],
\end{equation}
which is negligible unless $n=k=0$. Evaluation for this case shows
that (\ref{slipcoeff0}) has the more accurate form,
\begin{equation}\label{slipcoeff1}
\frac{\lambda}{\sqrt{DL}}\sim\frac{3}{16}\sqrt{\frac{\pi}{\phi_s}}
-\frac{1}{\pi}\left[\sqrt{\frac{L}{D}}\ln\left(\frac{D}{L}+
\sqrt{1+\frac{D^2}{L^2}}\right)+\frac{1}{2}\sqrt{\frac{D}{L}}
\ln\left(\frac{L}{D}+\sqrt{1+\frac{L^2}{D^2}}\right)\right].
\end{equation}
As a difference with the leading-order term, the next-order term in \eqref{slipcoeff1} is not symmetric in $(D,L)$, and therefore rectangular lattices display anisotropic friction. Note also that this next-order term 
 does not depend on the post radius $a$ (and therefore $\phi_s$). 
 For a square array ($L=D$),
(\ref{slipcoeff1}) simplifies to
\begin{equation}\label{slipcoeff1DD}
\frac{\lambda}{L}\sim\frac{3}{16}\sqrt{\frac{\pi}{\phi_s}}
-\frac{3}{2\pi}\ln(1+\sqrt{2}),
\end{equation}
in the limit of low $\phi_s$.
\section{Results and comparison with previous work}

\begin{figure}
\centering
\includegraphics[width=0.7\textwidth]{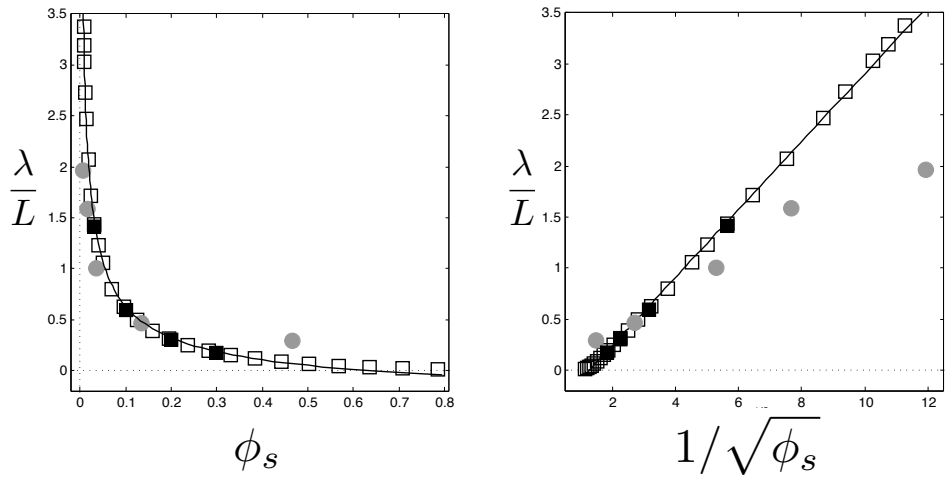}
\caption{Results of our model, and comparison with previous computational and experimental work, in the case of a square lattice. Effective slip length, $\lambda/L$, as a function of the solid surface fraction, $\phi_s$ (left), and as a function of  $1/\sqrt\phi_s$ (right).
Solid line: Simple asymptotic low $\phi_S$ model, \eqref{slipcoeff1DD};
Empty  squares: Computations  by \cite{ng09_2};
Filled squares: Computations by  \cite{ybert07};
Filled  circles: Experimental results by \cite{lee08}. 
}
\label{results}
\end{figure}

We now compare the results of our asymptotic model to previous work considering  circular posts organized on a square array. The comparison is displayed in figure~\ref{results}, where we show the effective slip length, $\lambda/L$, as a function of the solid surface fraction, $\phi_s$ (left) and as a function of  $1/\sqrt\phi_s$ (right). The solid line is our model, namely \eqref{slipcoeff1DD}. The symbols are data from previous work. Specifically, empty  squares are computations  by \cite{ng09_2}, filled squares are computations by  \cite{ybert07}, and filled  circles are experimental results by \cite{lee08}.

The quantitative agreement between our model and previous numerical work is remarkable. 
The square root dependance on $\phi_s$, evident in figure \ref{results} (right) is indeed reproduced by our model.  For the coefficient $A$ in \eqref{scaling} we predict here $A =3\sqrt\pi/16 \approx 0.332$.  
In their numerical simulations, \cite{ybert07} fit \eqref{scaling} to their results and obtain $A=0.325$, whereas \cite{ng09_2} fit to $A=0.34$, which are both within about two percent of our prediction.

For the second coefficient in \eqref{scaling}, we showed above  that a term independent of $\phi_s$  is indeed  the next-order term in the asymptotic expansion, and we obtained $B={3}\ln(1+\sqrt{2})/{2\pi}\approx 0.421$.  \cite{ybert07} fit $B=0.44$, which is  less than five percent larger, while  \cite{ng09_2} obtain $A=0.468$,  a value about ten percent above our result.
Overall, the integrated error between our simple model, \eqref{slipcoeff1DD}, and the  numerics from \cite{ng09_2} is about 1.8\%, while the  error between our model and the computations of \cite{ybert07} is about 3.9\%.

Our geometry implies that $\phi_s\leq \pi/4$ but the asymptotic estimate  \eqref{slipcoeff1DD}  loses validity at  lower values, as seen on the the right-hand-side of  figure \ref{results}(left), and  as illustrated by its prediction of no-slip ($\lambda =0$) at $\phi_s= \pi^3/[8\ln(1+\sqrt{2})]^2\approx 62\%$.

Regarding the comparison with the experiments of \cite{lee08}, we see that our model (and past computations) are able to predict the correct order of magnitude of the slip length, but do not capture the exact dependance on the surface friction, which appears to be weaker than inverse square  root (see figure~\ref{results}, right). This disagreement is likely due to the difference in flow field. In their paper, 
 \cite{lee08} measure the friction by using a cone-and-plate rheometer with rotation axis perpendicular to the surface place. As a result, the flow in their device has circular streamlines, and is not always aligned with the lattice periodicity. As a difference, in our work, as well as in \cite{ybert07} and \cite{ng09_2}, the flow considered is a planar shear flow  aligned with the lattice period.

\section{Conclusion}
In this work we considered theoretically shear flow past fakir-like super-hydrophobic surfaces composed of  circular posts located on a doubly-periodic rectangular lattice.  Using a superposition of point forces with suitably spatially-dependent strength, we obtained the total flow as an infinite series of linear equations. In the asymptotic limit of small surface coverage by the posts, the series can be interpreted as  Riemann sums,  which allowed us to  derive analytically the effective surface slip length in the form of \eqref{scaling}. In the case of a  square lattice, our results were found to be in excellent quantitative agreement with previous numerical computations. Future work will focus on embedding such low-friction fakir surface  on a curved substrate (for example, a sphere), Êand attempting to predict its mobility coefficient in a viscous fluid.

\begin{acknowledgments}
We thank Chiu-On Ng and Christophe Ybert for providing us the numerical results from \cite{ng09_2} and \cite{ybert07} respectively, as well as Choongyeop Lee and Chang-Jin ``CJ'' Kim  for providing us the experimental results from \cite{lee08}.  
We also thank Lyderic Bocquet for useful discussions. 
This work was supported in part by the National Science Foundation (Grant CBET-0746285 to E.L.).
\end{acknowledgments}

\bibliography{bioposts}

\end{document}